\documentclass[12pt,preprint]{aastex}           

   \newcommand{\tnm}{\tablenotemark}
   \newcommand{\gsim}{\rlap{$>$}{\lower 1.0ex\hbox{$\sim$}}}

   \renewcommand{\sec}{\prime\prime}
   
   \renewcommand{\deg}[0]{\circ}

   \shorttitle{Characterization of Extragalactic 24micron Sources}
   \shortauthors{Yan et al}

   \normalsize

\begin{document}

\title{Characterization of Extragalactic 24micron Sources in the Spitzer 
First Look Survey}

\author{Lin Yan\altaffilmark{1}, George Helou\altaffilmark{1},
D. Fadda\altaffilmark{1}, F.R. Marleau\altaffilmark{1}, M. Lacy\altaffilmark{1}, 
G. Wilson\altaffilmark{1}, B.T. Soifer\altaffilmark{1,2}, I. Drozdovsky\altaffilmark{1},
F. Masci\altaffilmark{1}, L. Armus\altaffilmark{1}, H.I. Teplitz\altaffilmark{1}, D.T. Frayer\altaffilmark{1}, J. Surace\altaffilmark{1}, L.J. Storrie-Lombardi\altaffilmark{1},
P.N. Appleton\altaffilmark{1}, S. Chapman\altaffilmark{2}, P. Choi\altaffilmark{1}, 
F. Fan\altaffilmark{1}, I. Heinrichsen\altaffilmark{1}, M. Im\altaffilmark{4}, 
M. Schmitz\altaffilmark{3}, D.L. Shupe\altaffilmark{1}, G.K. Squires\altaffilmark{1}}

\altaffiltext{1}{Spitzer Space Telescope Science Center, California
  Institute of Technology, 1200 East California Boulevard, MS 220-6,
  Pasadena, CA 91125; Send offprint requests to Lin Yan:
lyan@ipac.caltech.edu.}

\altaffiltext{2}{The Caltech Optical Observatories, Caltech, Pasadena,  CA~91125}
\altaffiltext{3}{IPAC, Caltech, Pasadena, CA 91125}
\altaffiltext{4}{School of Earth and Environmental Sciences, 
Seoul National University, Shillim-dong, Kwanak-gu,Seoul, S. Korea}

\begin{abstract}
In this Letter, we present the initial characterization of extragalactic 24$\mu m$
sources in the Spitzer First Look Survey (FLS) by examining their counterparts at 8$\mu m$ and $R$-band.
The color-color diagram of 24-to-8 vs. 24-to-0.7$\mu m$ is populated with 
$18,734$ sources brighter than the 3$\sigma$ flux limit of 110$\mu$Jy.
The data covers a total area of 3.7~sq.degrees. The 24-to-0.7$\mu m$ colors
of these sources span almost 4 orders of magnitudes, while the 24-to-8$\mu m$ colors
distribute at least over 2 orders of magnitudes. In addition to identifying
$\sim$30\%\ of the total sample with infrared quiescent, mostly low redshift galaxies, we also found that: (1) 
23\%\ of the 24$\mu m$ sources ($\sim$1200 per sq.degrees) with 
$\log_{10}(\nu f_{\nu}(24\mu m)/\nu f_{\nu}(8\mu m)) \ge 0.3$ and 
$\log_{10}(\nu f_{\nu}(24\mu m)/\nu f_{\nu}(0.7\mu m)) \ge 1.0$ are probably infrared luminous starburst galaxies
with $L_{IR} \ge 3 \times 10^{11} L_\odot$ at $z \ge 1$. In particular, 13\%\ of the sample
(660 per sq.degrees) are 24$\mu m$ detected only, with no detectable emission in either 8$\mu m$ or $R$-band.
With such extremely red IR/visible and mid-IR colors, these sources are the candidates for being
ULIGs at $z \ge 2$.  (2) 2\%\ of the sample (85 per sq.degrees) have extremely red mid-infrared to optical 
color ($\log_{10}(\nu f_{\nu}(24\mu m)/\nu f_{\nu}(0.7\mu m)) \ge 1.5$) and
fairly moderate 24-to-8$\mu m$ color ($\log_{10}(\nu f_{\nu}(24\mu m)/\nu f_{\nu}(8\mu m)) \sim 0.5$), 
and they are likely candidates for being dust reddened AGNs, like Mrk231 at $z \sim 0.6-3$. 
(3) We anticipate that some of these sources with extremely red colors may be new types of sources, since they
can not be modelled with any familiar type of spectral energy distribution.
We find that close to 38\%\ of the 24$\mu m$ sources have optical $R$ fainter than 23.0 vega magnitude,
and 17\%\ of these have no detectable optical counterparts brighter than $R$ limit of 25.5mag.
Optical spectroscopy of these optical extremely faint 24$\mu m$ sources would be very difficult, and
mid-infrared spectroscopy from the Spitzer would be critical for understanding their physical nature.

\end{abstract}

\section{Introduction}

The far-IR background detected by {\it COBE}
(Puget et al.\ 1996; Fixsen et
al.\ 1998) peaks around $\sim200\mu$m with energy comparable to
the optical/UV background. 
This implies that ~50\%\ of the integrated rest-frame optical/UV
emission is thermally reprocessed by dust and radiated at mid to far-infrared.
Thus, dust enshrouded galaxies with  high IR/visible ratios, particularly
Ultraluminous Infrared Galaxies 
(ULIRGs, $L_{fir} > 10^{12}$~L$_\odot$, $L_{fir}/L_{vis} > $a few),
make a significant contribution to the total energy budget and star formation
over the history of the Universe.
Deep surveys based on the rest-frame UV/optical do not provide a complete
census of galaxy populations, and their measurements of
luminosity and SFR are only lower limits (Meurer, Heckman \&\ Calzetti 1999; Yan et al. 1999).
Deep ISO and sub-mm SCUBA observations have shown that the
integrated luminosity density from dusty sources peak around $z \sim 1$, 
roughly a factor of 4 - 5 higher than those measured from
the rest-frame optical surveys (Serjeant et al. 2000; Elbaz et al. 1999;
Gruppioni et al. 2003; Blain et al. 1999). This peak in the SFR appears to be relatively flat out
to $z \sim 2 - 4$ (Franceschini et al. 2001; Franceschini et al. 2002; Elbaz et al. 2002; Lagache et al. 2003).
 
While ISO deep surveys probe primarily galaxies at $z < 1$ (Genzel \&\ Cesarsky 2000),
and sub-mm SCUBA observations are limited to a small number of high luminosity dusty 
sources at $z > 2$ (Chapman et al. 2003), the dusty universe at $ z > 1$ is largely unexplored
by systematic surveys in the mid-to-far infrared wavelength.
The 24$\mu m$ imaging camera on the Spitzer Space Telescope (Werner et al. 2004) 
provides us the first opportunity to do this. IRAS and ISO studies have 
shown that mid-infrared emission is a good indicator of the bolometric IR luminosity
(Soifer et al. 1987; Surace et al. 1998; Elbaz et al. 2002; Chary \&\ Elbaz 2001).
At $ 0.7 < z < 2.5$, the 24$\mu m$ band samples redshifted, rest-frame 
6--12$\mu$m emission from polycyclic aromatic hydrocarbons (PAH)
and very small dust grains in dusty galaxies, making it the most sensitive window to
probe the high-redshift infrared bright galaxies.

In this Letter, we make the first attempt to characterize the properties
of 24$\mu m$ selected galaxies within the Spitzer First Look Survey (FLS)\footnote{see http://
ssc.spitzer.caltech.edu/fls for details}.
This paper studies the distribution in the 
24-to-8 and 24-to-0.7micron (R(24,8) and R(24,0.7)\footnote{here we define R(24,8) 
$\equiv \log_{10}(\nu f_{\nu}(24\mu m)/\nu f_{\nu}(8\mu m))$, and R(24,0.7) $\equiv \log_{10}(
\nu f_{\nu}(24\mu m)/\nu f_{\nu}(0.7\mu m)$)} 
color-color diagram
of 18,734 24$\mu m$ selected sources over an area of 3.7~square degree.
The R(24,8) and R(24,0.7) colors are indicative of the intrinsic slopes
of the spectral energy distribution (SED) as well as dust extinction
and K corrections.  This Letter pays special attention to the populations with 
extremely red 24-to-8 and 24-to-0.7$\mu m$ colors, and estimate
their surface densities. Throughout this paper, we adopt $H_0=70$km/s/Mpc, $\Omega_{M}=0.3$
and $\Omega_{\Lambda} = 0.7$. The magnitude system is in vega.

\section{Observations and Data Reduction}

\subsection{Image Reduction and Source Extraction}

Table 1 summarizes all of the data went into the analyses in this Letter.
Specifically, the 24$\mu m$ flux cutoff of 110$\mu$Jy is for SNR of 3$\sigma$, which
is computed at the peak pixel. This is a conservative measurement in comparison with the
SNR calculated through a large aperture. After the reliable sources were selected with this conservative
flux limit, we can use lower SNR (2$\sigma$) to estimate the flux limits for 
sources not detected in the 8$\mu m$ and $R$ band. The Spitzer data were taken with the IRAC and 
MIPS cameras (Fazio et al. 2004; Rieke et al. 2004). The 8$\mu m$ and 24$\mu m$ data cover an area of 
3.7~square degrees, where these two datasets overlap. 
Both IRAC and MIPS raw data were processed and mosaiced together by the pipeline 
provided by the Spitzer Science Center (SSC). Additional corrections
to the Basic Calibrated Data (BCD)
images were included, see Lacy et al. (2004) and Fadda et al. (2004)
for details. The 8$\mu m$ source catalog is generated using SExtractor (Bertin \&\ Arnout 1996).
We used the photometry within a $6.1^{''}$ diameter aperture. 
The total fluxes are obtained by applying the appropriate aperture corrections, which
are scaled to the $24^{''}$ diameter aperture. This is at most around 30\% (Lacy et al. 2004).
The 24$\mu m$ source catalog was generated using StarFinder (Diolaiti et al. 2000).
The aperture correction for obtaining the total 24$\mu m$ flux is $\sim$10\%
(Fadda et al. 2004a).
The R-band images were taken with the MOSAIC-1 camera on the 4~m telescope at the
Kitt Peak National Observatory. The reduced and stacked images as well as source catalogs have been publicly 
released, and  the detailed description of the observations,
data reduction and catalogs can be found in Fadda et al. (2004b).

\subsection{Bandmerged Catalog}

We cross-identified the 24$\mu m$ sources in the 8$\mu m$ and $R$-band 
by simple positional matching.
The IRAC astrometry was fine-tuned using the reference positions from
the 2MASS point sources. The MIPS 24$\mu m$ images were aligned with 
the VLA 20cm radio positions. The mean positional differences between $R$ and 
24$\mu m$ for bright and unsaturated sources are 0.1$^{''}$ with a rms of 0.4$^{''}$ in both
right ascension and declination. Similarly, we
found the $<\Delta RA> = 0.04^{''}\pm 0.6^{''}$ and $<\Delta DEC> =
0.03^{''}\pm0.6^{''}$ between the IRAC 8$\mu m$ and $R$-band images.
We use $2^{''}$ matching radii, generously chosen to allow 
somewhat larger astrometric uncertainties for fainter sources. 
Multiple matches between the 24 and 8 $\mu m$ catalogs are
neglible, less than 0.05\%. The fraction of multiple
matches between the 24 micron and R-band catalogs is higher,
about 7\%. In these cases, we choose the closest matches 
in the positional centroids between the two bands.
Stellar contamination for a 24$\mu m$ selected sample 
is expected to be low since we are sampling the tail of the Rayleigh-Jeans 
energy distribution and the FLS galactic latitute is 37$^\deg$.
Stars brighter than $R$ of 20~mag can be easily
identified on the $R$-band images. 
For sources with $ 20 < R < 23$, we used the stellarity index ($> 0.8$ sources are stars)
measured with SExtractor. The total stellar contamination 
is estimated to be only 7.5\%. We have visually spot-checked the final catalog on the image display.

\section{Results}

We have a total of 18,734 sources, which
are detected at 24$\mu m$ with fluxes brighter than $3\sigma$ 110$\mu$Jy.
Of these, close to half of them have good detections in all three bands.
The remaining half of the 24$\mu m$ sources have no detections in one or both of 
the 8$\mu m$ and $R$ bands. 
Plate 1 gives the 2-dimensional stamp images of a set of representative examples. 
As shown, some bright 24$\mu m$
sources have no counterparts at either 8$\mu m$ or $R$ band. These cases will be discussed
in detail below.

\begin{figure*}
\plotone{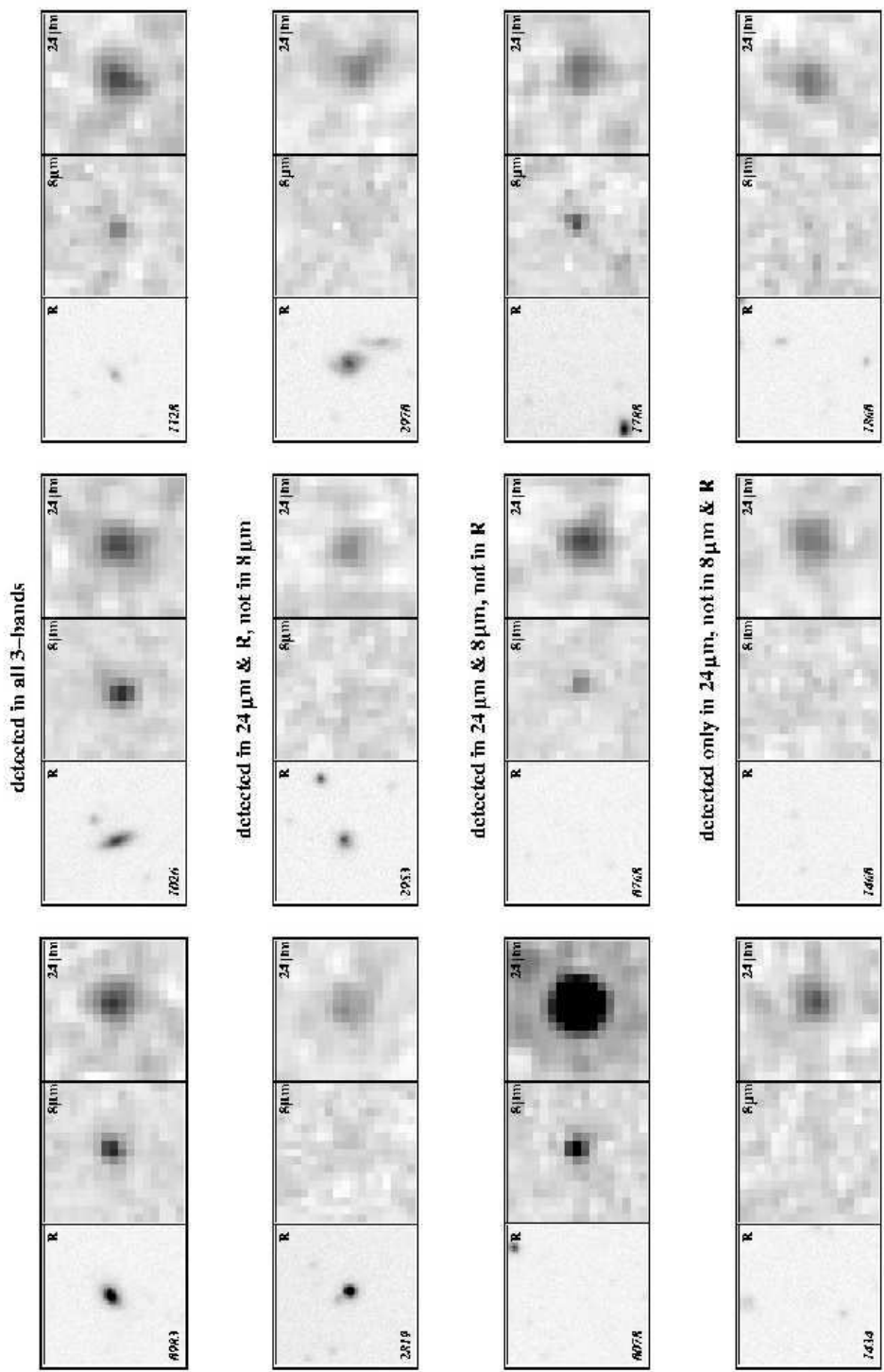}
\caption{This figure is also included as fig1.jpg.
{\bf Plate 1} shows a few representative examples of four types of 24$\mu m$ sources:
sources detected in all three bands; sources detected in 8$\mu m$ but not in $R$; 
sources detected in $R$ but not in 8$\mu m$; 24$\mu m$ detected sources only. }
\end{figure*}
 
The main result of this Letter is presented in Figure 2a, showing
the 24-to-8 and 24-to-0.7$\mu m$ color-color distribution of our sources.
Figure 2b presents the expected color-color tracks as a function
of redshifts computed from known types of SED templates. Figure 2a shows a broad correlation
between higher IR/visible ratios and IR colors, indicative of more intensely
heated dust, similar to the trends seen in the IRAS data (Soifer \&\ Neugebauer 1991).
Using this figure, we can
crudely classify various types of 24$\mu m$ selected sources by comparison with
the expected colors of known SED templates as well as with 
known objects in the FLS region. 
The SED templates include Arp220 and M82 (Silva et al. 1989; Chary \&\ Elbaz 2001),
NGC253 (Fadda et al. 2002), radio-quiet QSO, M51 \&\ NGC1068 (Dale et al. 2001; 
Dale \&\ Helou 2002).
To show the colors of an early type galaxy,
we construct the SED for the bulge of M31 within an aperture of $4^{'}$ diameter
using the near-IR and IRAS data published in Soifer et al. (1986). The optical
part of the SED for M31 bulge was taken as a 10~Gyr old elliptical SED from
Bruzual \&\ Charlot\footnote{ftp://gemini.tuc.noao.edu/pub/charlot/bca5}, 
then scaled and matched with
the infrared part to produce a full SED covering from 0.1 --- 100~$\mu$m.

\begin{figure}[tbh!]
\plotone{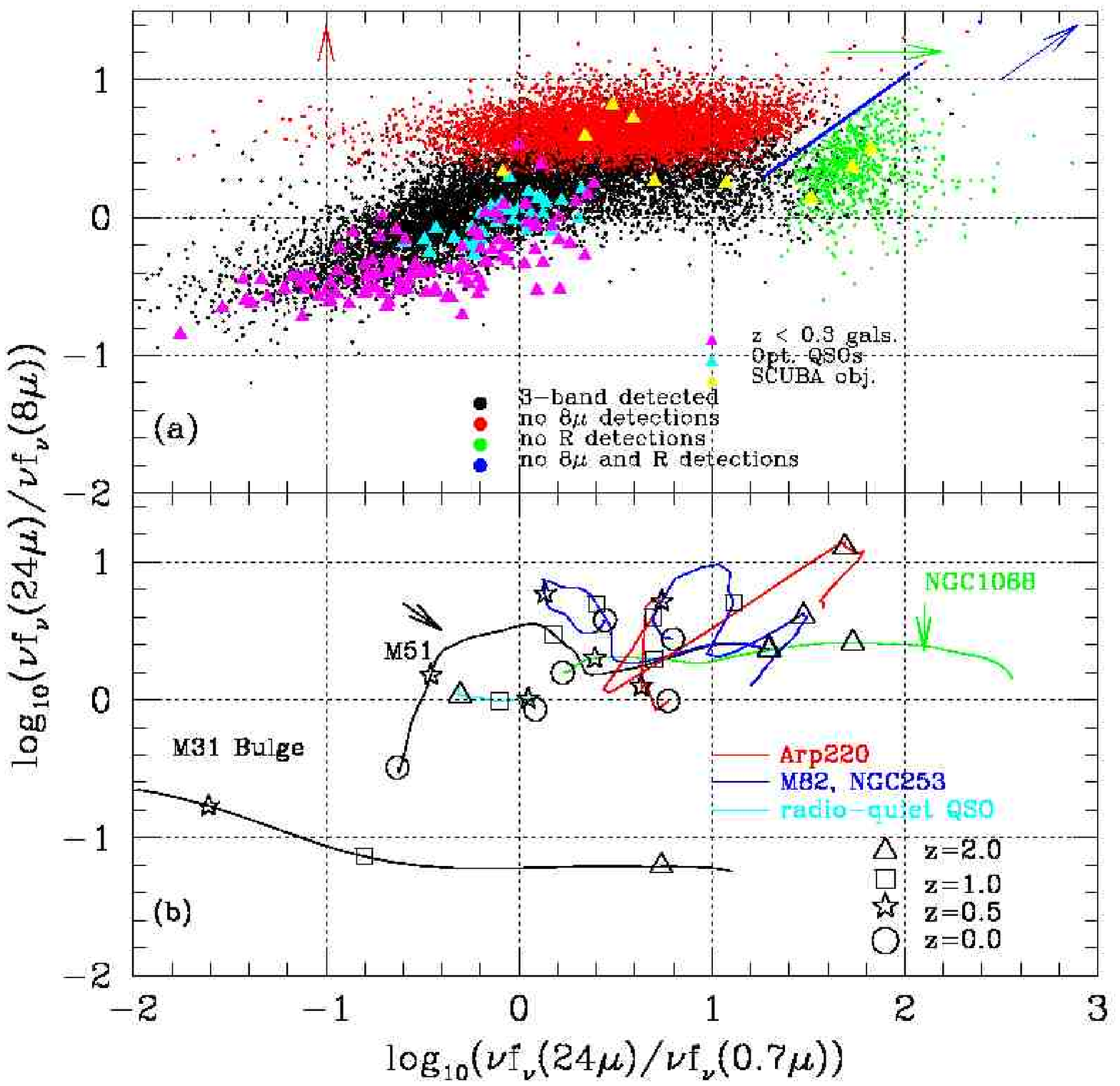}
\caption{This figure is also included as fig2.jpg.
The color-color plot between $\log_{10}(\nu f_{\nu}(24\mu)/\nu f_{\nu}(8\mu))$ 
and $\log_{10}(\nu f_{\nu}(24\mu)/\nu f_{\nu}(R))$. {\bf Panel a} shows the data
for a total of 18,734 sources selected at 24$\mu m$ in the FLS main field.
The small black points are sources detected in
all three bands; the  small green points are sources only detected
in the 8$\mu m$ bands, and not detected in the optical R-band;
the small red points indicate sources
with significant detections in the R-band, but no detections at 8$\mu m$;
the small blue points represents the sources detected only in the 24$\mu m$.
To illustrate the lower limits in colors,
we marked the arrows with corresponding color code for the red, green and blue sources.
The big yellow triangles are sources with SCUBA detections, the cyan triangles
are optically selected QSOs with known redshifts and spectra; the magenta triangles
are low redshift galaxies identified from the SDSS. {\bf Panel b} shows the expected
color tracks as a function of redshifts computed from a set of SED templates for 
local known galaxies.
}
\end{figure}

\subsection{Infrared Quiescent Sources}

Comparing Figure 2a and 2b, the color-color distribution can be crudely
classified into two extreme regions --- the infrared quiescent region where $R(24,8) \le 0.5$ and $R(24,0.7) \le 0$,
the infrared luminous region with $R(24,8) > 0.5$ and $R(24,0.7) > 1$. The region
between these two probably contains a mixture of starbursts and normal galaxies at various redshifts.
In Figure 2a, the black points occupy the most of the first region.
These are 24$\mu m$ sources which have significant detections at 8$\mu m$ as well as $R$-band.
Comparing with the model color-color tracks, we infer that the sources with $R(24,8) \le 0.5$ and $R(24,0.7) \le 0$ 
are primarily normal, infrared faint galaxies at low redshift ($z < 0.7$) and
optically selected QSOs with a wider redshift distribution.
Figure 3a,b show the differential number counts as functions of 
colors, $R(24,8)$ and $R(24,0.7)$. A little less than 30\%\ of the total 
population are infrared quiescent spirals and early type galaxies at low redshifts as well
as optically selected QSOs. The galaxies with extremely low 24-to-8micron flux ratio are old bulges, such as M31. 
Their colors can not be explained by pure interstellar dust emission, and they must have substantial
stellar photospheric and dust envelope contributions coming into
the 8$\mu m$ band.

\begin{figure}[h!]
\plotone{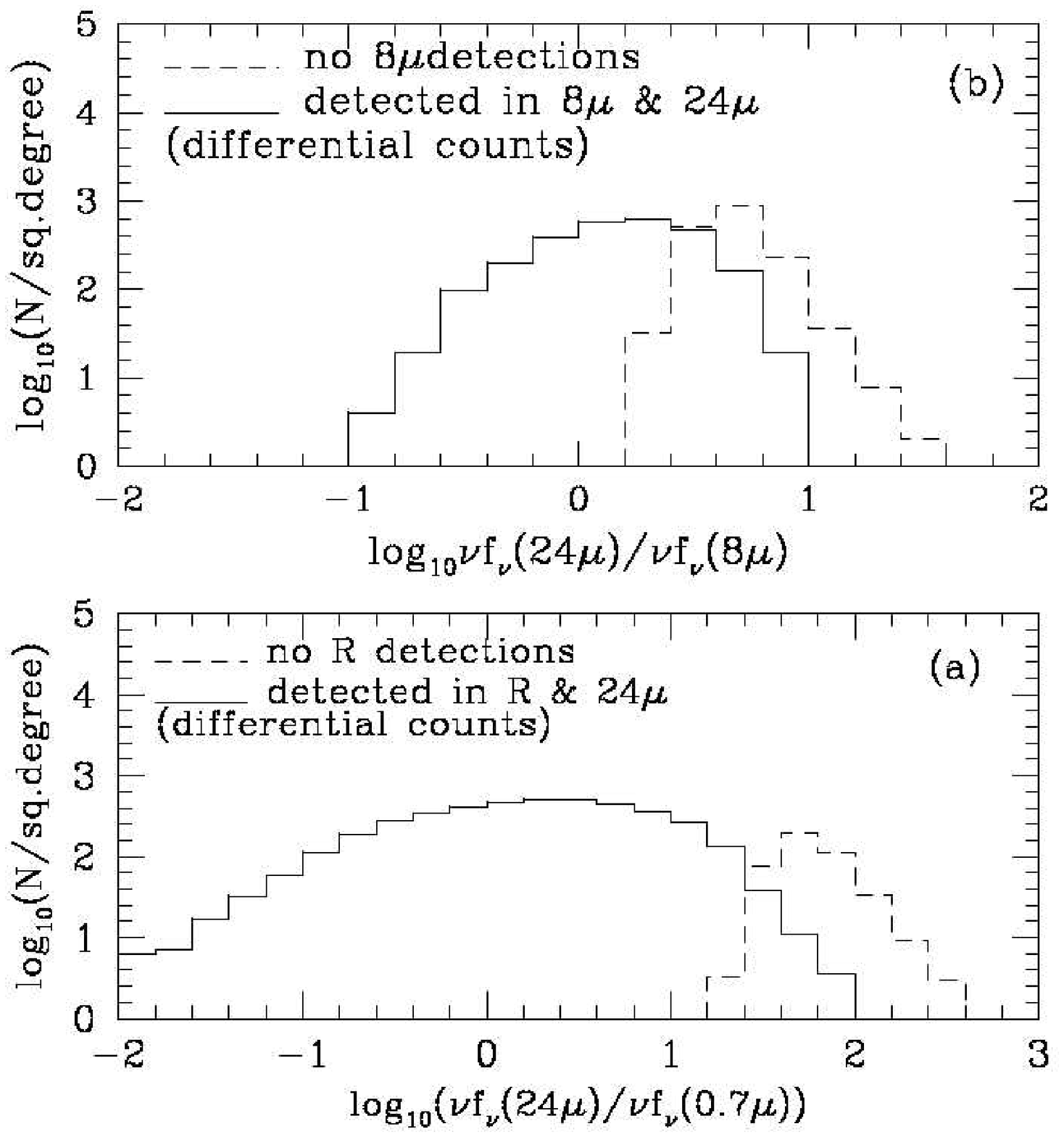}
\caption{This figure is also included as fig3.jpg.
The 24$\mu m$ source counts as functions of $R(24,0.7)$ and
$R(24,8)$ The dashed lines in both panels
are sources with flux limits in either the 8$\mu m$ or the R-band. The number counts
are differential, and the y-axis has units of number of sources per square degree.
The total area coverage is 3.7 sq.degree, and the total number of sources is 18,734.}
\end{figure}

To test further this interpretation of Figure 2a, 
we have used the NED extragalactic database
and identified the sources in our catalog with the
low redshift galaxies (magenta triangles) 
and optically selected QSOs (cyan triangles) from
the Sloan Digital Sky Survey{\footnote{http://www.sdss.org}}.
The locations of these sources with known types and redshifts
in Figure 2a generally confirm our interpretation.
The cyan points in Figure 2a can be well explained by the computed colors
of a radio-quiet QSO SED template in Figure 2b.
These optically selected QSOs have fairly flat SEDs
(Neugebauer et al. 1987; Sanders et al. 1989),
therefore, their colors are around $R(24,8) \sim 0$ and $R(24,0.7) \sim 0$
with very little change with redshifts.
One specific example is a sloan QSO at $z = 4.5445$, with the fluxes of 744$\mu$Jy,
98~$\mu$Jy and 27~$\mu$Jy at the 24, 8 and 0.7$\mu$m respectively, thus
$R(24,8)=0.4$ and $R(24,0.7)=-0.1$.

\subsection{The Nature of the Extreme 24$\mu m$ Populations}

Figure 2a illustrates the colors of four types of sources marked
with small black, red, green and blue points.
The objects indicated with black points have been discussed before.
The red points consist 31\%\ of the 24$\mu m$ sources, which are detected only in $R$-band, but 
not at the 8$\mu m$-band. The green points represent 4\%, with detections only in 8$\mu m$, 
but no counterparts in $R$ brighter than 2$\sigma$ limit of 25.5mag.
Finally, 13\%\ of the sample (660 objects per sq.degree) are not detected in either 
8$\mu m$ or $R$, shown as small blue points in Figure 2a. 
The mean fluxes of the objects marked as red, green and blue points 
are roughly between 230-360$\mu$Jy.
One of the most interesting sources revealed by the 24$\mu m$ images are ones which have
$R(24,0.7) > 1$ and $R(24,8) > 0.3$ in Figure 2a. These sources 
are candidates for being luminous starburst galaxies
at $z \ge 1$. Using this color-color cut,
we estimate that 23\%\ ($\sim$1200 per sq.degree) of our 24$\mu m$ sample are such objects.  
In particular, those sources detected only at 24$\mu m$ (blue points)
are likely to be ULIGs at $z > 2$
(660 per sq.degrees).  At $z = 1$, our 3$\sigma$ 24$\mu m$ flux 
limit of 110$\mu$Jy corresponds to the observed luminosity $\nu L_{\nu}$ 
of $2\times 10^{10} L_\odot$.  If we ignore the filter difference between the Spitzer
24$\mu m$ and the IRAS 12$\mu m$ (the 24$\mu m$ bandwidth is about half of the 12$\mu m$ filter), 
the $z=1$ observed 24$\mu m$ luminosity is roughly
the rest-frame IRAS 12$\mu m$ luminosity. The total infrared luminosity
$L_{IR}(8 - 1000\mu)$ is correlated with IRAS 12$\mu$m luminosity
$\nu L_{\nu}(12\mu)$, in the form of $L_{IR} = 0.89\times 
\nu L_{\nu}(12\mu)^{1.094} L_\odot$, 
as derived from the IRAS Revised Bright Galaxy Sample (Soifer et al. 1987; Chary \&\ Elbaz 2001).
This implies that the $z > 1$ starbursts should have $L_{IR} \ge 2.7\times 10^{11} L_\odot$.
As suggested by the 24$\mu m$ source counts (Marleau et al. 2004),
the FLS data may reach $z \sim 2$ starburst population.
Our flux limit implies that at $z \sim 2$, the infrared luminosity $L_{IR}$
should be brighter than $2\times 10^{12} L_\odot$.

Another interesting population are sources 
with $R(24, 0.7) > 1.5$ but with a fairly constant $R(24,8) \sim 0.5$. 
Figure 2b suggests that these sources are likely to be dust heavily reddened
AGNs like NGC~1068, or Mrk231 at $z > 1$. These sources 
could be separated from the general population using the IRAC 8-to-4.5 and 5.8-to-3.5micron 
color-color selection (Lacy et al. 2004).
The surface density of these dusty AGNs are on the order of $\sim$85 per sq.degree, and they 
constitute $\sim$2\%\ of the total 24$\mu m$ population.
To confirm our prediction that we can use $R(24,8)$ vs. $R(24,0.70$ to select
high redshift, infrared luminous galaxies, we mark on Figure 2a the several 
24$\mu m$ sources detected with the SCUBA at 
850$\mu m$ (see Frayer et al. 2004 for detail). 
These SCUBA sources could be at $z \sim 1 - 3$, as shown in Chapman et al. (2003).
Our computed color-color tracks in Figure 2 is to illustrate 
in a broad sense what range of colors each type of sources should have.
A small number of sources with extremely red 24-to-8 and 24-to-R colors 
can not be all explained by models with known type of SEDs. This could be suggestive
of potentially new classes of objects.


We examine the optical
brightness of 24$\mu m$ selected sources in Figure 4, showing
the $R(24,0.7)$ and $R(24,8)$ as a function
of $R$ magnitude for the 24$\mu m$ sources. Close to 36\%\
of the 24$\mu m$ sources have $R$ magnitudes fainter than 23.0mag.
Such a faint optical magnitude  might suggest that they are at $z > 0.5$, supported by
the measurements from optical $R$ band selected redshift surveys (Lilly et al. 1996;
Cohen et al. 2001).  Many of these sources are well within the brightness regime of 
the 10~meter classes telescopes for obtaing the optical spectroscopic redshifts. 
However, the 24$\mu m$ sources without $R$ counterparts fainter than 2$\sigma$ of 25.5mag
will be difficult to follow-up with the optical spectroscopy. The redshifts and 
physical natures of these sources could be measured using 
the mid-infrared spectrograph (IRS) from the Spitzer InfraRed Spectrometer. 
Particularly, for sources with 24$\mu m$ fluxes brighter than 750$\mu$Jy, 
the IRS would be sensitive enough
to obtain good SNR low resolution spectra covering 5$\mu m$ to 38$\mu m$ 
with a total of (1-2)~hours of integration. We have a total of 46 sources 
which have 24$\mu m$ fluxes
brighter than 750$\mu$Jy and $R$-band magnitude fainter than 25.0mag.

\begin{figure}[h!]
\plotone{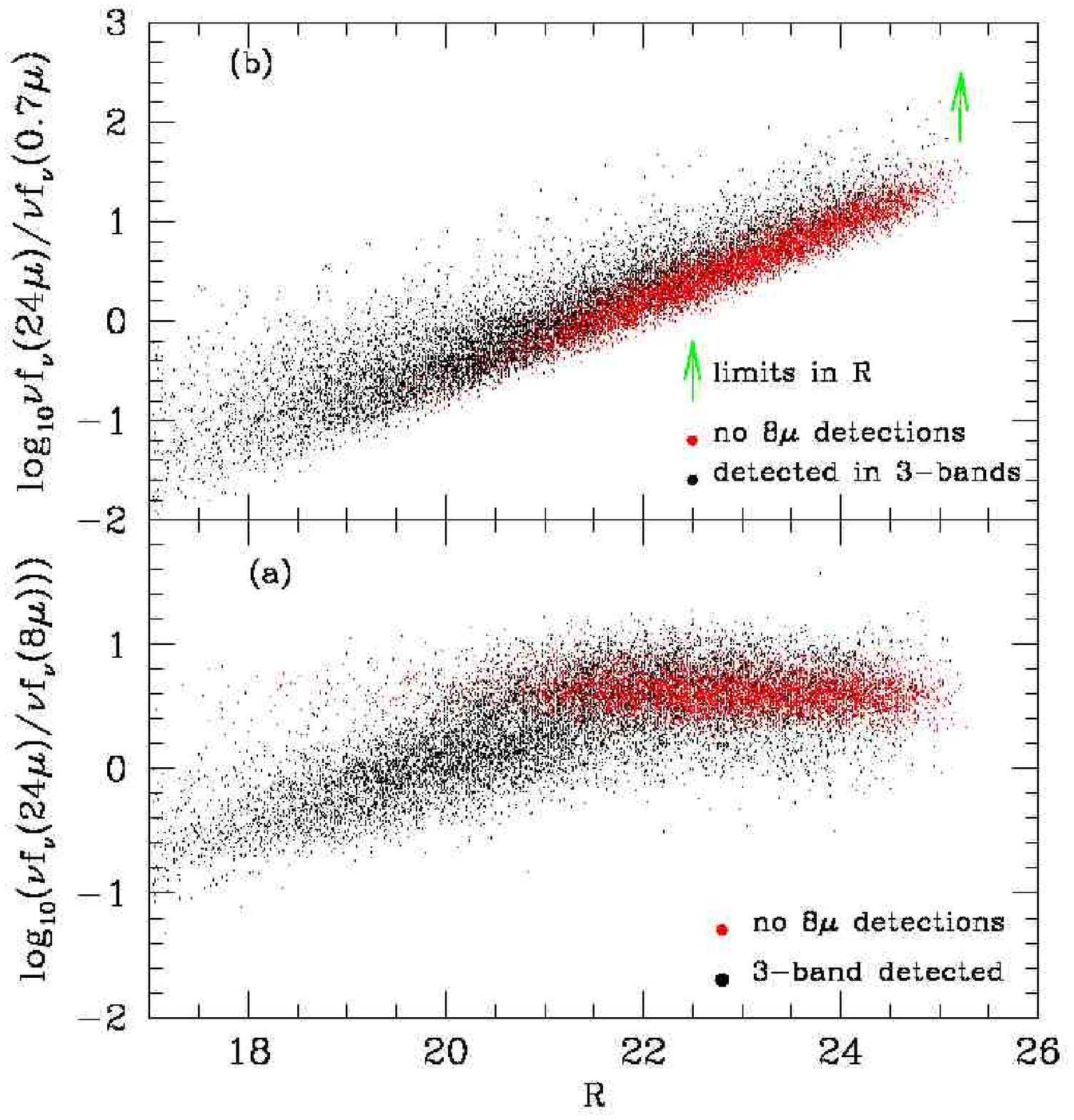}
\caption{This figure is also included as fig4.jpg.
{\bf Panel a} shows the 24-to-8$\mu m$ flux ratio versus 
$R$ magnitude for the 24$\mu m$ selected sample. 
{\bf Panel b} shows the color-magnitude 
diagram of the 24-to-0.7$\mu m$ flux ratios versus $R$ magnitude. }
\end{figure}

This work is based in part on observations made with the Spitzer Space Telescope, 
which is operated by the Jet Propulsion Laboratory, California Institute of Technology 
under NASA contract 1407. Support for this work was provided by NASA."
We also made use of the NASA/IPAC Extragalactic Database (NED) which is operated
by the Jet Propulsion Laboratory, California Institute of Technology, under
contract with the National Aeronautics and Space Administration.

\clearpage

\begin{deluxetable}{lccc}
   \tablewidth{0pt}
   \tabletypesize{\scriptsize}
   \tablecaption{Summary of Optical, IRAC 8$\mu m$ and MIPS 24$\mu m$ Observations}
\startdata
\hline\hline \\
$\lambda_{cent}$ ($\mu$m) & 0.7 & 8.0 & 24.0 \\
FWHM ($\sec$)          & 1.0 & 2.2 & 5.5 \\
Flux Limits & 25.5mag\tnm{a} & 20$\mu$Jy\tnm{a} & 110$\mu$Jy\tnm{a} \\
$\nu f_{\nu}(lim)$(ergs/s/cm$^{2}$) & 1.3e-15\tnm{b} & 3.75e-15 & 24e-15 \\
$<$Exp. Time$>$ (sec) & 1800 & 60 & 80 \\
\enddata
\tablenotetext{a}{Here $R$ of 25.5mag is a 2$\sigma$ limit within a $3^{''}$ diameter aperture. 
The 8$\mu m$ flux 20$\mu$Jy is a 2$\sigma$ limit, included aperture correction. The 24$\mu m$ flux
110$\mu$Jy is a flux cutoff where the SNR at the peak pixel is greater than or equal to 3$\sigma$.
See the text for details.}
\tablenotetext{b}{$f_{R}(0mag)=2780$Jy (Neugebauer 1997, private communication)}
\end{deluxetable}


\end{document}